\documentclass[12pt, draftclsnofoot, onecolumn]{IEEEtran} 
 \usepackage{amsmath,amssymb}
 \usepackage{subfigure}
 \usepackage{graphicx,graphics,color,psfrag}
 \usepackage{cite,balance}
 \usepackage{algorithm}
 \usepackage{accents}
 \usepackage{amsthm}
 \usepackage{bm}
 \usepackage{url}
 \usepackage{algorithmic}
 \usepackage[english]{babel}
 \usepackage{multirow}
 \usepackage{enumerate}
 \usepackage{cases}
 \usepackage{stfloats}
 \usepackage{dsfont}
 \usepackage{color,soul}
 \usepackage{amsfonts}
  \usepackage{tcolorbox}
\usepackage[utf8]{inputenc}
 \usepackage{cite,graphicx,amsmath,amssymb}
 \usepackage{subfigure}
 \usepackage{fancyhdr}
 \usepackage{hhline}
 \usepackage{graphicx,graphics}
 \usepackage{array,color}
 \usepackage{amsmath}
 \usepackage{amsthm}

\include{header}

\IEEEoverridecommandlockouts

\begin{document}

\title{An Overview on Over-the-Air Federated Edge Learning}
\author{Xiaowen Cao, Zhonghao Lyu, Guangxu Zhu,  Jie Xu, Lexi Xu, and Shuguang Cui
\thanks{X. Cao, Z. Lyu, and J. Xu are with the School of Science and Engineering (SSE) and the Future Network of Intelligence Institute (FNii), The Chinese University of Hong Kong (Shenzhen), Shenzhen 518172, China. (e-mail: caoxwen@outlook.com, zhonghaolyu@link.cuhk.edu.cn, xujie@cuhk.edu.cn).}
\thanks{G. Zhu is with Shenzhen Research Institute of Big Data, Shenzhen 518172, China  (e-mail: gxzhu@sribd.cn). G. Zhu and J. Xu are corresponding authors. }
\thanks{ L. Xu is with Research Institute, China United Network Communications Corporation, Beijing, China (email: xulx29@chinaunicom.cn).}
\thanks{S. Cui is with the SSE and FNii, The Chinese University of Hong Kong (Shenzhen), and Shenzhen Research Institute of Big Data, Shenzhen 518172, China. He is also affiliated with Peng Cheng Laboratory, Shenzhen, China, 518066 (e-mail: shuguangcui@cuhk.edu.cn).}
}

\markboth{}{}
\maketitle

\setlength\abovedisplayskip{2pt}
\setlength\belowdisplayskip{2pt}

\vspace{-1.5cm}

\begin{abstract}
\emph{Over-the-air federated edge learning} (Air-FEEL) has emerged as a promising solution to support edge \emph{artificial intelligence} (AI) in future \emph{beyond 5G} (B5G) and 6G networks. In Air-FEEL, distributed edge devices use their local data to collaboratively train AI models while preserving data privacy, in which the over-the-air model/gradient aggregation is exploited for enhancing the learning efficiency. This article provides an overview on the state of the art of Air-FEEL. First, we present the basic principle of Air-FEEL, and introduce the technical challenges for Air-FEEL design due to the over-the-air aggregation errors, as well as the resource and data heterogeneities at edge devices. Next, we present the fundamental performance metrics for Air-FEEL, and review resource management solutions and design considerations for enhancing the Air-FEEL performance. Finally, several interesting research directions are pointed out to motivate future work.

\end{abstract}

%

\section{Introduction}\label{sec:intro}


Recently, edge learning or edge intelligence has been recognized as a promising technique towards {\it beyond 5G} (B5G) and 6G to enable emerging applications such as meta-verse, smart city, automated driving, and industrial automation, in which {\it artificial intelligence} (AI) is relocated from the central cloud to the network edge to allow prompt data access and exploitation with significantly reduced latency. Among a variety of edge learning paradigms, {\it federated edge learning} (FEEL) is particularly appealing, due to its advantages in efficient model training and data privacy preservation by exploiting the distributed computing and storage capabilities at network edge \cite{MChenJSAC_ML}. Essentially, FEEL is implemented based on the distributed {\it stochastic gradient descent} (SGD) over wireless networks, which aims to find optimized AI-model parameters by minimizing a properly designed loss function in an iterative manner. To be specific, at each global iteration or communication round, the edge server first broadcasts the global model to edge devices for synchronizing their local models; next, the edge devices update their respective local models by performing one or more local gradient descent iterations; finally, the edge devices upload their local gradients or models to the edge server for aggregation, thus yielding an updated global model  to initiate the next-round training. 
Based on the distributed SGD principle, FEEL can be implemented via two typical approaches, namely the {\it federated stochastic gradient descent} (FedSGD) and {\it federated averaging} (FedAvg), in which edge devices upload the local gradients and models after one and multiple local iterations, respectively, at each communication round \cite{FedAvg}. However, due to the frequent exchange of high-dimensional models or gradients (usually comprising millions to billions of parameters) between the edge server and edge devices, the communication cost becomes the paramount performance bottleneck for FEEL, especially when the number of participating edge devices becomes large.

   \begin{figure}
\centering
 \setlength{\abovecaptionskip}{-1mm}
\setlength{\belowcaptionskip}{-1mm}
    \includegraphics[width=6.5in]{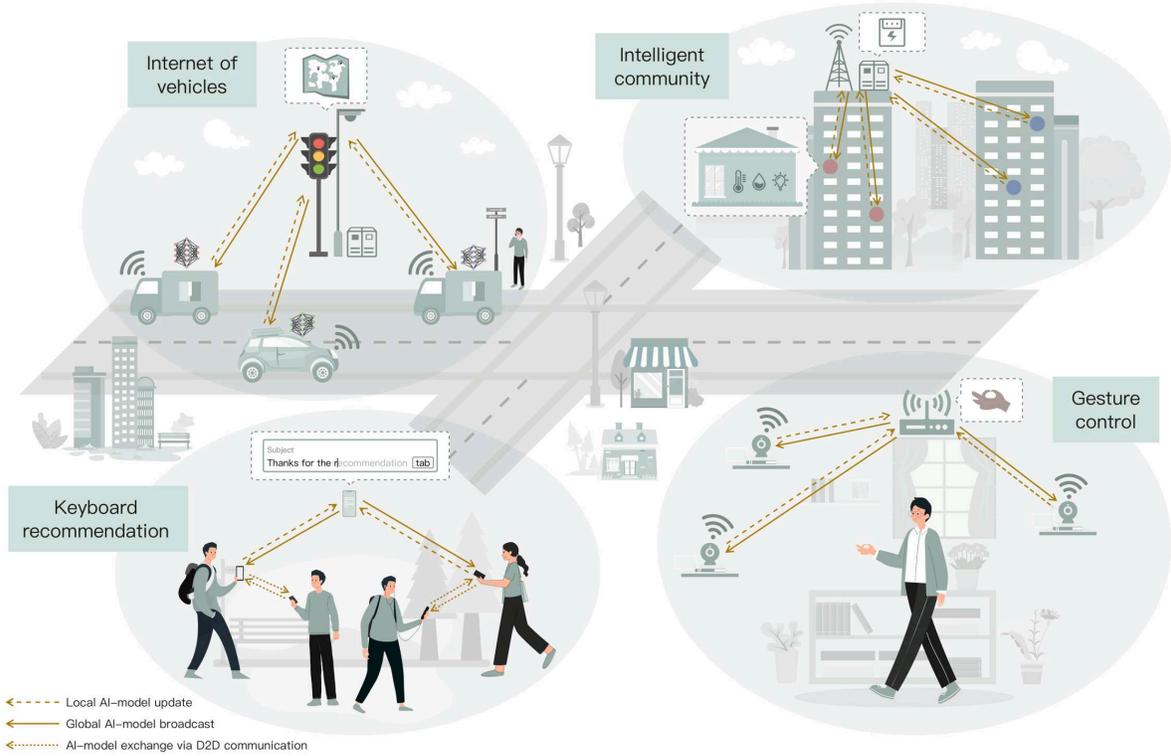}
\caption{Example application scenarios of Air-FEEL. } \label{fig:Application}
\vspace{-0.1cm}
\end{figure}


To tackle this issue, {\it over-the-air FEEL} (Air-FEEL) has been proposed recently as a promising solution, in which distributed edge devices are allowed to concurrently transmit their local gradient or model updates for ``one-shot'' aggregation at each communication round.
In such a way, the communication and computation are seamlessly integrated to enhance the communication efficiency in FEEL, thus supporting various low-latency and communication-efficient applications, such as Internet of vehicles and intelligent community, as shown in Fig. \ref{fig:Application}. 
The magic behind the promising Air-FEEL applications is the so-called {\it over-the-air computation} (AirComp) technique, which utilizes the superposition property of a {\it multiple-access channel} (MAC) for distributed functional computation over the air \cite{Zhu2021ComMag}. 
As such, Air-FEEL can achieve remarkable communication latency reduction up to a factor equal to the device population, as compared to the conventional FEEL based on the classic communicate-then-compute approach for gradient/model aggregation. 
Moreover, Air-FEEL is able to further enhance the data privacy by combating the potential eavesdropping attack during the gradient/model uploading process. This is due to the fact that with AirComp, potential eavesdroppers can only access to the aggregated updates with each local component hidden in the crowd \cite{DLiu2020Ar}. 

Due to the aforementioned advantages, Air-FEEL is becoming a hot research topic in the field of edge intelligence, which has attracted growing research interests from both academia and industry. However, the successful implementation of Air-FEEL to materialize its promising gain faces various technical challenges resulting from the over-the-air aggregation errors as well as the resource and data heterogeneities at distributed edge devices. Various research efforts have been conducted in the literature to tackle these issues. 

In view of the growing research interests, this article provides an overview on the state of the art of Air-FEEL. First, we introduce the basic principle of Air-FEEL, and discuss its fundamental performance metrics in terms of training loss and learning accuracy, learning latency, and energy consumption. Next, we review various  solutions to enhance the Air-FEEL performance, covering power control, beamforming, user scheduling, and large-scale network optimization. Finally, we point out several interesting research directions to motivate future work on Air-FEEL.

\section{Basic Principle of Air-FEEL}

\begin{figure}
\centering
 \setlength{\abovecaptionskip}{-1mm}
\setlength{\belowcaptionskip}{-1mm}
    \includegraphics[width=5.6in]{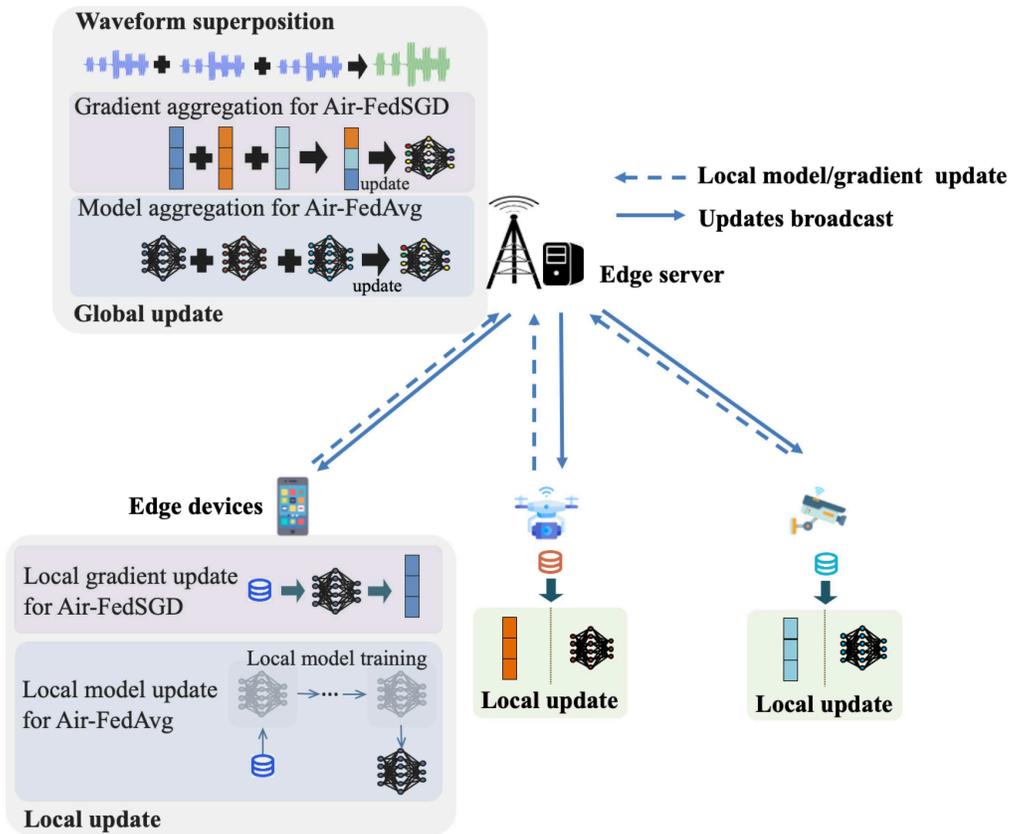}
\caption{Basic architecture of Air-FEEL. } \label{fig:model}
\vspace{-0.1cm}
\end{figure}

This section introduces the basics of Air-FEEL. A typical Air-FEEL system is shown in Fig. \ref{fig:model}, where one edge server coordinates multiple distributed edge devices to train an AI model based on the ``one-shot'' over-the-air model/gradient aggregation. 
As mentioned earlier in the introduction, {\it over-the-air FedSGD} (Air-FedSGD) and {\it over-the-air FedAvg} (Air-FedAvg) are two typical approaches for Air-FEEL implementation.
In particular, at each global communication round, Air-FedSGD employs the over-the-air {\it gradient} aggregation after {\it one} local computation iteration, while Air-FedAvg performs the over-the-air {\it model} aggregation after {\it multiple} local computation iterations. 
Generally, Air-FedSGD enjoys better robustness to the non-{\it independent and identically distributed} (i.i.d.) data, while Air-FedAvg requires much less frequent model aggregation and thus lower communication cost.

The main difference of Air-FEEL from other FEEL paradigms is that, Air-FEEL exploits AirComp for one-shot over-the-air model/gradient aggregation \cite{Zhu2021ComMag}, as explained in the following.
For example, consider the basic setup with one edge server and $K$ edge devices in an {\it additive white Gaussian noise} (AWGN) channel. Consider a particular global communication round, in which each device $k$ needs to upload its local gradient/model parameter ${\bf w}_k \in {\mathbb C}^{N \times 1}$ with $N$ dimensions, and the desired global parameter at the edge server is $\frac{1}{K}\sum_{k=1}^K {\bf w}_k$.
In this case, after acquiring the information of its own channel coefficient $h_k$, each device $k$ multiplies ${\bf w}_k$ by the pre-processing coefficient $\alpha_k \frac{h_k^c}{| h_k|}$ to amplify the signal and compensate the channel phase, and then  transmits $\alpha_k \frac{h_k^c}{|h_k|} {\bf w}_k$ over $N$ channel uses, where superscript $c$ denotes the conjugate, $|\cdot|$ denotes the absolute value of a complex number, and $\alpha_k$ denotes the transmit amplitude depending on the transmit power. Accordingly, the received signal at the edge server becomes ${\bf y} = \sum_{k=1}^K \alpha_k|h_k| {\bf w}_k + {\bf z}$ with a noise vector ${\bf z}$ over $N$ channel uses at edge server receiver. Then, after post-processing by multiplying denoising factor $1/\eta$ and averaging operation\footnote{With proper pre- and post-processing, AirComp can go beyond averaging to compute a series of nomographic functions such as geometric mean, weighted sum, polynomial, and Euclidean norm \cite{Zhu2021ComMag}.}, we have the AirComped  parameters as ${\bf r}=\frac{1}{\eta K}{\bf y}=\sum_{k=1}^K \frac{\alpha_k|h_k| {\bf w}_k}{\eta K} +\frac{ {\bf z}}{\eta K}$. 
Especially, the {\it aggregation error} caused by AirComp with respect to the desired average parameter $\frac{1}{K}\sum_{k=1}^K {\bf w}_k$ is ${\bm \varepsilon} = {\bf r}-\frac{1}{K}\sum_{k=1}^K {\bf w}_k$.
For instance, when $h_{k}=h$ and $\alpha_{k}=\sqrt{P}$, we can set $\eta = h\sqrt{P}$ to get ${\bf r}=\frac{1}{K}\sum_{k=1}^K {\bf w}_k+\frac{ {\bf z}}{Kh\sqrt{P}}$ with a noise-induced error ${\bm \varepsilon} = \frac{ {\bf z}}{Kh\sqrt{P}}$. 
It is clear that by properly designing $\{ \alpha_k\}$, we are able to obtain the desired average value of gradient or model parameters for FEEL, but subject to certain distortion caused by {\it amplitude misalignment} (i.e., when $\alpha_k|h_k|$'s are different over edge devices) and noise ${\bf z}$. 

\begin{figure}
\centering
 \setlength{\abovecaptionskip}{-1mm}
\setlength{\belowcaptionskip}{-1mm}
    \includegraphics[width=6.8in]{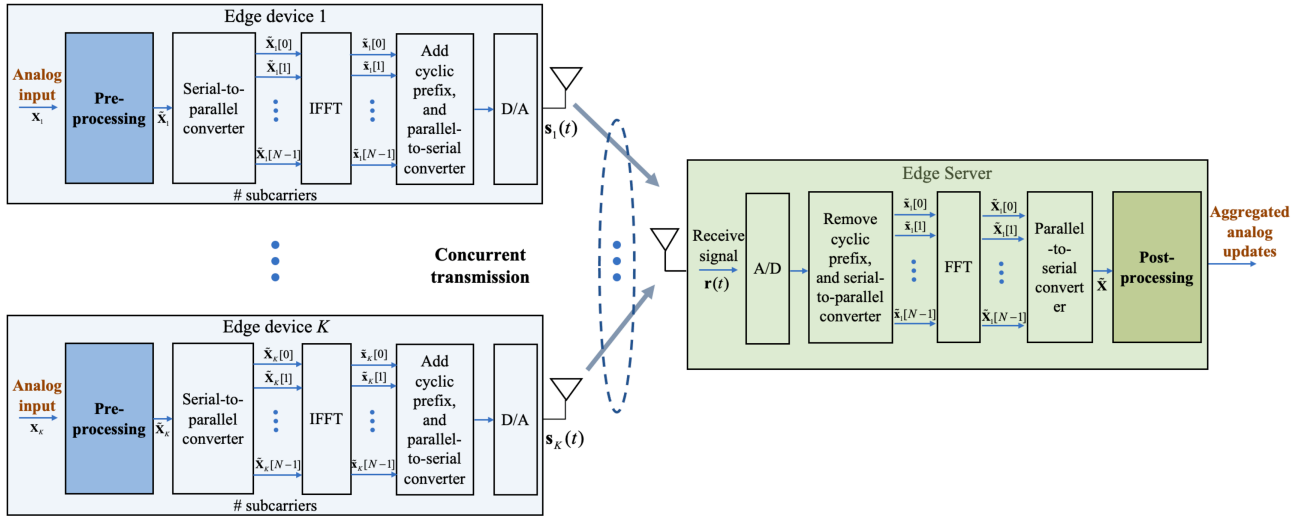}
\caption{Illustration of wideband digital AirComp implementation over OFDM. } \label{fig:OFDM}
\vspace{-0.1cm}
\end{figure}

Notice that AirComp and Air-FEEL designs can also be implemented over fading wideband channels, in which the above process can be implemented over each subchannel independently. 
For instance, Fig. \ref{fig:OFDM} shows the digital realization of AirComp over an {\it orthogonal frequency multiplex access} (OFDM) system, which is different from conventional OFDM communications in the following two aspects. First, we need to add an additional  pre-processing at each edge device to compensate the channel fading and phase shift, and proper post-processing at the edge server receiver. Next, analog input is implemented for AirComp in Fig.~\ref{fig:OFDM}, which is in sharp contrast to the  conventional OFDM communications using digital bits as inputs.
For more details on AirComp, please refer to the overview paper \cite{Zhu2021ComMag}. 

\subsection{Benefits of Air-FEEL}

The benefits of Air-FEEL are two-fold. First, Air-FEEL is advantageous in enhancing the communication efficiency and reducing the training latency. In particular, in Air-FEEL, the over-the-air aggregation of each AI-model parameter from $K$ edge devices only requires one single channel use, thus making the required total number of resource blocks and the training latency independent from $K$. This is in sharp contrast to the conventional design with sequential ``communication-then-computation'', in which the edge server needs to decode individual AI-model parameters from $K$ devices by mitigating the inter-user interference, such that the required total number of resource blocks and the training latency generally grow with respect to $K$ linearly.

Next, Air-FEEL is beneficial in better preserving data privacy. In conventional FEEL, raw data exposure to the third party is avoided by exchanging model/gradient updates; however, private data information can still be inferred from the individual updates via membership inference or update leakage attacks \cite{DLiu2020Ar}, thus causing certain level of privacy leakage. By contrast, in Air-FEEL, only the aggregate updates, instead of individual ones, can be accessed by potential eavesdroppers, thus providing double privacy protection at edge devices. 



\subsection{Challenges}\label{Challenges}
The successful implementation of Air-FEEL faces the following technical challenges.

\subsubsection{Over-the-air Aggregation Errors}

In Air-FEEL, analog magnitude modulation is implemented for one-shot distributed functional computation. This, however, is vulnerable to signal distortion caused by channel fading, noise, and channel estimation errors, and thus may degrade the training performance. How to quantify the effect of such over-the-air aggregation errors on Air-FEEL performance and accordingly optimize the system design is an uncharted problem that has not been investigated in conventional FEEL and wireless communication systems.


%

\subsubsection{Resource Limitation and Heterogeneity}\label{Resource}

 While the distributed learning tasks are generally resource-consuming, the participating edge devices only have limited computation and communication resources, and their available resources may vary significantly. Such resource limitation and heterogeneity are becoming a challenging issue faced in Air-FEEL. On one hand, the over-the-air gradient/model aggregation needs to be implemented synchronously by all edge devices after their local iteration, but their local iteration latencies may differ significantly due to their distinct computation capabilities. In this case, the total computation latency of Air-FEEL is restrained by the device with the smallest computation power. On the other hand, the over-the-air aggregation errors highly depend on the channel conditions and the transmit power at distributed edge devices. In this case, devices with poor channel and/or limited transmit power become the performance bottleneck of AirComp, and increasing their transmit powers can accelerate the Air-FEEL training but at a cost of high energy consumption cost.

\subsubsection{Data Heterogeneity}\label{DataResource}
Data heterogeneity is another key factor affecting the performance of Air-FEEL. Specifically, in practical wireless networks, training data are distributed at different edge devices in a non-i.i.d. manner, and the number of available data samples may be highly unbalanced among different edge devices. The non-i.i.d. and unbalanced data distribution poses a grand challenge in designing good Air-FEEL systems with convergence and generalization guarantee. 


\section{Fundamental Performance Limits }\label{Performance}


To facilitate the Air-FEEL design, it is important to understand its fundamental performance limits. This section first presents the widely adopted performance metrics in Air-FEEL, and then discusses their fundamental tradeoffs under practical over-the-air aggregation errors.

\subsection{Performance Metrics}
Air-FEEL aims to train AI models. The widely adopted performance metrics include training loss and learning accuracy, learning latency, and energy consumption, which are detailed in the following.

\begin{itemize}
\item {}  {\bf Training loss and learning accuracy}: Training loss means the loss (or objective) function value of the trained AI model, and learning accuracy describes how the trained model performs over test samples. In general, with a given AI model, lower training loss corresponds to higher learning accuracy. To analyze the convergence behavior of Air-FEEL, loss optimality gap (the gap between the current loss and the global minimum) is widely adopted in practice  \cite{DLiu2020Ar,Cao2021-FedAvg,Cao2021AirFEEL}.

\item {}   {\bf Learning latency}: Learning latency is defined as the wall clock time for the learning process to converge within a given accuracy. In Air-FEEL, the learning latency generally corresponds to the number of global communication rounds times per-round latency \cite{Cao2021-FedAvg}. Here, the per-round latency includes the computation latency (of the slowest device) for local gradient/model-update iterations, and the communication latency for over-the-air updates aggregation from the edge devices to the edge server.

\item {}  {\bf Energy consumption}: Energy consumption corresponds to the communication and computation energy required by both the edge devices and edge server for training the AI model. The computation energy is mainly consumed by the edge devices for locally updating AI models, and the communication energy is consumed by the edge devices and edge server to exchange their models/gradients.

\end{itemize}

Intuitively, the above three Air-FEEL performance objectives are competing, e.g., to achieve higher accuracy normally requires longer learning latency and more energy consumption. Such performance tradeoffs, however, are quite complicated and depend on various system design parameters, such as the numbers of global and local iterations, the transmit power at edge devices for AirComp, and their computation power for local computation. In the literature, there have been well-established analytic models on the learning latency and energy consumption for Air-FEEL \cite{Zhu2021ComMag}. It remains a fundamental problem to analytically understand the relationship between the training loss and learning accuracy versus system parameters under different AI models, especially subject to the unique over-the-air aggregation errors.


\subsection{Convergence of Air-FEEL with Aggregation Errors }\label{ConvergenceAna}

To resolve the above issue, it is essential to analyze the convergence behavior of Air-FEEL, which characterizes how the (expected) gradient norm or loss optimality gap changes over global communication rounds. However, it is difficult to find the exact expression for gradient norm or loss optimality gap under general AI models. To tackle this problem, their upper bounds are alternatively used to represent the convergence behavior in the FEEL literature \cite{DLiu2020Ar}, by making specific assumptions on the AI model (e.g., on the smoothness of the loss functions and Lipschitz-continuous gradients). However, due to the newly involved over-the-air aggregation errors, the convergence analysis for Air-FEEL is still challenging.

\begin{figure}
\centering
 \setlength{\abovecaptionskip}{-1mm}
\setlength{\belowcaptionskip}{-1mm}
    \includegraphics[width=6in]{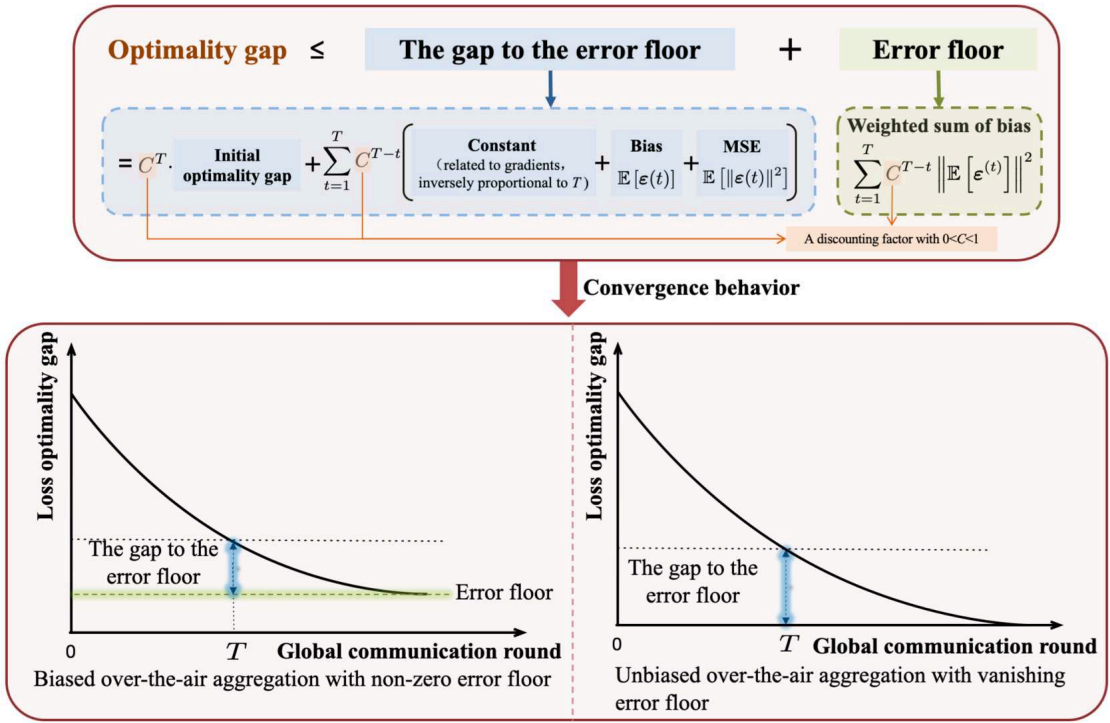}
\caption{Illustration of the convergence behavior by taking into account over-the-air aggregation errors \cite{Cao2021AirFEEL}, where $T$ denotes the needed global communication rounds for convergence with $1\leq t\leq T$, and $\mathbb{E}\left[{\bm \varepsilon}(t)\right] $ and $\mathbb{E}\left[\|{\bm \varepsilon}(t)\|^{2}\right]$ denote the bias and MSE of the aggregation error at each global communication round $t$, respectively. } \label{fig:contration}
\vspace{-0.1cm}
\end{figure}

Recently, the work \cite{Cao2021AirFEEL} established a convergence analysis for Air-FedSGD with gradient aggregation after one local iteration. This framework reveals the impact of the bias and MSE of aggregation errors on the excepted optimality gap. In particular, under the smoothness assumption and Polyak-{\L}ojasiewicz inequality, the loss optimality gap is shown to satisfy the simplified inequality as shown in the upper frame in Fig.~\ref{fig:contration}.

The lower frame in Fig.~\ref{fig:contration} illustrates the convergence behavior of Air-FEEL in terms of the derived loss optimality gap under the specified assumptions \cite{Cao2021AirFEEL}.
 First, the Air-FEEL algorithm converges towards the optimality within a certain error floor, which becomes zero if the over-the-air aggregation is unbiased but strictly positive otherwise. 
 Nevertheless, enforcing unbiased aggregation for vanishing error floor may elevate the gap to the error floor, and thus may slow down the convergence. Next, with the increasing discounting factor, the aggregation errors at the latter global rounds contribute more to the optimality gap and thus are more damaging to the learning process.

The above analytical result in \cite{Cao2021AirFEEL}  sheds light on how the imperfect aggregate updates affect the convergence of Air-FedSGD. This result was then extended to Air-FedAvg with model aggregation after multiple local iterations \cite{Cao2021-FedAvg}. It is shown that, as the number of local iterations increases, the needed number of global communication rounds may decrease, but at the cost of a large divergence among the local updates at edge devices. This thus shows the significance of judiciously designing the number of local iterations in Air-FedAvg for improving the learning performance. These analytical results build the foundation for optimizing the Air-FEEL performance, as will be elaborated in the sequel.

\section{Resource Management for Air-FEEL}
Based on the performance analysis in Section \ref{Performance}, this section reviews the state-of-the-art techniques for  improving the Air-FEEL performance, which jointly optimizes the communication parameters for AirComp and the learning parameters.

\subsection{Power Control Optimization}

In Air-FEEL, the edge devices can adaptively control their transmit power to reduce the aggregation error for model/gradient aggregation, and accordingly enhance the learning accuracy or convergence rate. In general, there are two different transmit power control design principles. The first design principle optimizes  AirComp independently at each global round for, e.g., minimizing the communication error or MSE. For instance, the most straightforward way is for the edge devices to adopt the channel inversion power control for achieving the perfect signal magnitude alignment. This, however, is highly suboptimal in MSE minimization. This is because in this case, the aligned signal magnitude is compromised by the device with the worst channel conditions, which may significantly weaken the aggregate signal strength and equivalently amplify the noise power, especially when some devices are in deep fading. To deal with this issue, the optimal power control to minimize the AirComp MSE was developed in \cite{Zhu2021ComMag,NZhang2020Ar}, which exhibits a threshold-based structure depending on the channel conditions at different edge devices. For a certain device, if the channel power gain is greater than a certain threshold, then a channel-inversion-like power control is employed at that device for signal magnitude alignment; otherwise, the full power transmission is performed at that device to mitigate the noise.

\begin{figure}
\centering
 \setlength{\abovecaptionskip}{-1mm}
\setlength{\belowcaptionskip}{-1mm}
    \includegraphics[width=4in]{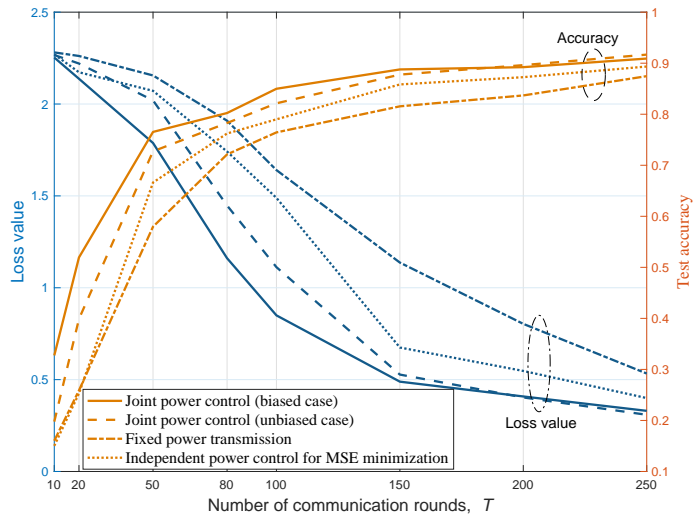}
  \caption{Learning performance of Air-FEEL on MNIST dataset over the number of communication rounds under different power control schemes \cite{Cao2021AirFEEL}.}
  \label{CNN_Fig:N}
\vspace{-0.1cm}
\end{figure}


Instead of minimizing the aggregation MSE in each round independently, the second design principle jointly optimizes the power allocation over different global rounds to maximize the convergence rate in terms of the loss optimality gap (e.g., in the upper frame in Fig.~\ref{fig:contration}). For instance, the work \cite{Cao2021AirFEEL} investigated the power allocation problem in Air-FedSGD, in which two joint power control designs with biased and unbiased aggregations are developed, respectively. 
As shown in Fig. \ref{CNN_Fig:N}, both joint power control designs achieve faster convergence and lower loss value compared with the fixed power transmission and independent power control for MSE minimization. This thus shows the benefit of power control optimization in accelerating the learning convergence. As compared with independent power control for MSE minimization,  the joint power control designs capture the fact that the loss optimality gap is more sensitive to the aggregation errors at latter rounds. 
Fig. \ref{CNN_Fig:N} also compares the loss value and learning accuracy of the two joint power control designs. It is shown that the design under the unbiased aggregation achieves lower optimality gap than that under the biased aggregation when the number of communication rounds is sufficiently large. This coincides with the discussion in Section~\ref{ConvergenceAna} that the Air-FedSGD algorithm converges to the optimal point with unbiased gradient aggregation.
The convergence-aware joint power control design was then extended to the Air-FedAvg system in \cite{Cao2021-FedAvg}.


\subsection{Beamforming Design}

Multi-antenna beamforming is an efficient solution to further improve the Air-FEEL performance by enhancing both aggregation efficiency and aggregation accuracy, via exploiting the spatial multiplexing and spatial diversity gains, respectively. On one hand, by deploying multiple antennas at both the edge devices and the edge server, multiple AI-model parameters or gradient elements can be simultaneously aggregated over one single time-frequency block. On the other hand, via proper transmit and/or receive beamforming, multiple signal copies from different antennas can be properly combined to further reduce the aggregated errors. For instance, existing works in \cite{Zhu2021ComMag,KYang2020TWC} developed the joint design of transmit beamforming at the edge devices and aggregation beamforming at edge server for substantially enhancing the aggregation performance, which can be directly used for enhancing Air-FEEL performance.

More recently, {\it reconfigurable intelligent surface} (RIS) has emerged as a candidate technique to enhance the spectrum and energy efficiencies of 6G networks \cite{WNi_IoT}. RIS is composed of a large number of low-cost and passive reflecting elements, whose amplitudes and phases can be adjusted to properly control the propagation environment to facilitate wireless transmissions. In particular, RIS can assist the edge devices with poor channel conditions to enhance their signal strengths towards the edge server, thus enhancing the aggregated signal magnitude level for minimizing the aggregation error. For RIS-empowered Air-FEEL, it is critical to jointly control the phase shifts of reflecting elements at RIS, the transmit beamforming at edge devices, and the receive beamforming at edge server for enhancing the learning performance \cite{WNi_IoT}.


\subsection{User Selection/Scheduling}

Use selection and scheduling is another efficient technique to enhance the Air-FEEL performance by resolving the resource heterogeneity issue via dropping the edge devices with bad channels and/or limited communication and computation resources.
A new tradeoff arises in scheduling edge devices for optimizing Air-FEEL performance.   In particular, allowing more devices to participate in Air-FEEL can aggregate more training data for accelerating the convergence, which, however, may lead to larger aggregation errors as more devices with poorer channels are involved. Such tradeoffs were first studied in \cite{GZhu2020TWC} to regulate the device population in the scenario when the edge devices are i.i.d. distributed in a certain area, in which the devices with weak channel conditions were excluded. Next, the joint design of device scheduling and receive beamforming was presented in \cite{KYang2020TWC}, in which the devices with weak signal strengths after receive beamforming were dropped from the training process. Moreover, the work  \cite{Sun_2021Sced} investigated the device scheduling by further considering their diverse energy constraints and computation capabilities, in which an energy-aware dynamic device scheduling algorithm was proposed based on Lyapunov optimization.


\subsection{Model/Gradient Compression Based on Sparsification}

Compression is useful in reducing the data dimensions to be exchanged, and thus is important to accelerate the communication process in Air-FEEL. However, different from FEEL that can apply conventional digital compression techniques, Air-FEEL needs to implement new compression methods based on analog modulated model/gradient parameters. For example, the authors in \cite{Amiri2020TSP} exploited the sparsity of the gradient/model updates to compress the local gradient estimates to the dimension of available channel uses.
In this design, edge devices first sparsify the gradient estimates by setting elements with small absolute values to zero, and then projected the processed gradient estimates into a low-dimensional vector by applying a synchronized pseudo-random measurement matrix. This thus avoids the additional transmission of index information of zero elements in gradient vector. 
Furthermore, the work \cite{JZhang2021Ar} considered gradient sparsification together with an error-feedback mechanism, in which  the common dimensions can be selected uniformly via synchronized pseudo-random number generators and the sparsification errors are added back to the next-round gradient update to alleviate the possible gradient error. Under this setup, the adaptive power control was implemented to combat the aggregation errors over time-varying channels.

\section{Future Research Directions}


Besides the above resource management designs, there also remain various new design issues that are important for the success of Air-FEEL in 6G, but not well investigated yet. In this section, we discuss some of these issues to motivate future work.

\subsection{Secure Design Against Jamming and Spoofing}

The analog over-the-air aggregation in Air-FEEL makes it vulnerable to jamming or spoofing attacks due to the lack of coding and encryption. For instance, some corrupted devices due to malfunction or malicious adversary (a.k.a. Byzantine devices) may send fault or manipulated messages to the edge server to attack Air-FEEL for preventing the convergence or leading to a counterfeit model. In this case, how to detect or even correct the manipulated abnormal model/gradient updates by malicious devices (i.e., poisoning attacks) on global model training is important. This task, however, is particularly challenging, due to the simultaneous model/gradient transmission over the air. 

To tackle the above issue, the authors in \cite{Sifaou2021Ar} proposed to use the Weiszfeld algorithm to obtain the smoothed geometric median of model/gradient for resisting the Byzantine attack. In this design, the edge devices are divided into several groups, based on which the devices within each group can employ AirComp for computing the geometric median of their local models/gradients, but those in different groups need to transmit over orthogonal time-frequency blocks. Based on such an over-the-air geometric median aggregation scheme, it is shown that when the number of Byzantine attacks is lower than half of the number of groups or individual transmissions, the aggregation values are ensured to achieve a well approximation to the mean of updates of honest devices. In essence, the design in  \cite{Sifaou2021Ar} combats against the Byzantine attacks at the cost of more communication costs. How to further improve such design and study new approaches to combat other attacks still requires more efforts from our community.


\subsection{Air-FEEL over Large-Scale Networks}

\begin{figure}
\centering
 \setlength{\abovecaptionskip}{-1mm}
\setlength{\belowcaptionskip}{-1mm}
    \includegraphics[width=6.5in]{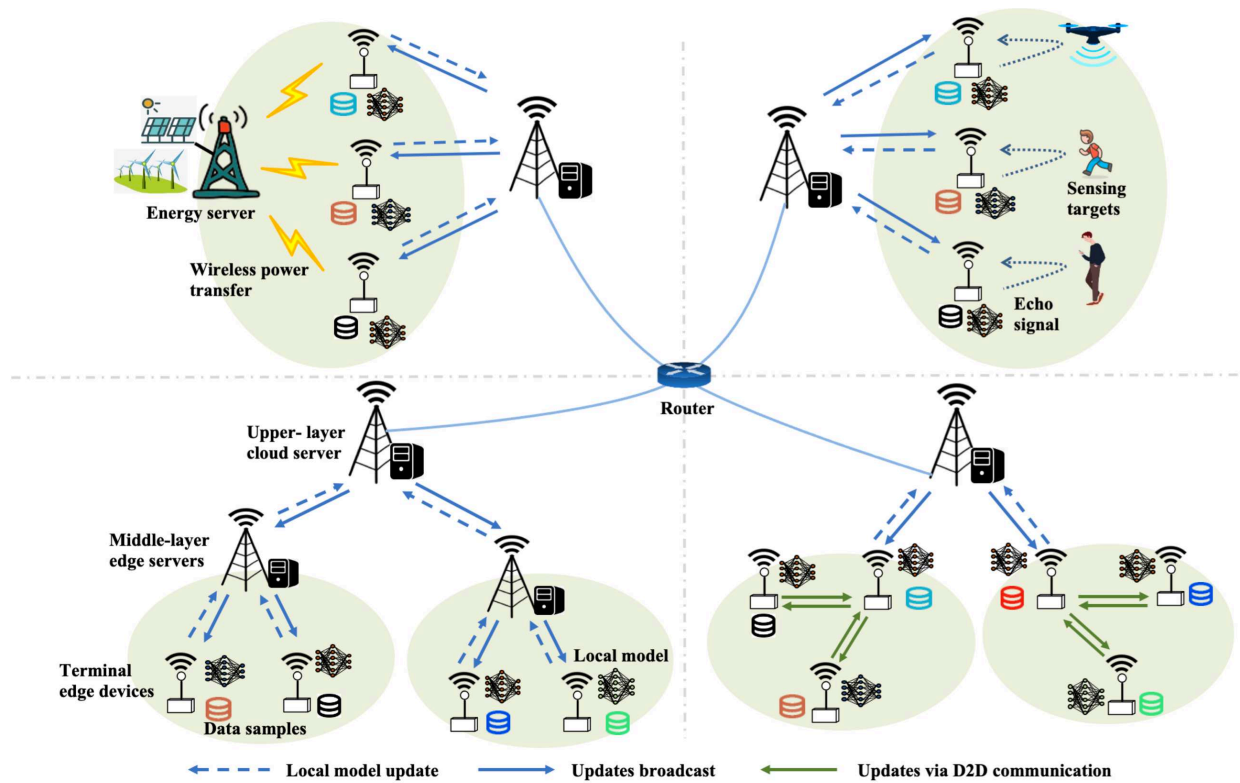}
\caption{Future Air-FEEL framework with potential solutions. } \label{fig:Hierachical}
\vspace{-0.1cm}
\end{figure}

The success of FEEL needs massive training data. It is thus desirable to involve more edge devices over large-scale networks for Air-FEEL. Towards this end, hierarchical Air-FEEL and {\it device-to-device} (D2D)  Air-FEEL can be two viable architectures, as depicted in the two lower sub-figures in Fig. \ref{fig:Hierachical}. In particular, the hierarchical Air-FEEL architecture may correspond to a multi-cell network with multiple edge servers and one (or more) cloud server at the core network. At the lower level, the edge server may aggregate local models/gradients from edge devices over the air; at the upper level, the cloud server performs global aggregation of regional updates from edge servers via wired or wireless backhauls. How to associate edge devices with edge servers, and how to determine the numbers of local computation iterations at the lower and upper levels are new problemsto tackle. These problems, however, are difficult, especially when the distributed data samples follow varied distribution at edge devices.

On the other hand, in the D2D Air-FEEL without a centralized controller, all participating edge devices are divided into multiple clusters, such that each device is associated with a cluster head  via D2D communications \cite{HXing2021Ar}. In this system, edge devices within the same cluster first simultaneously send the updates to the associated cluster head, and then the head broadcasts the aggregated (cluster) signal to all devices to achieve consensus. In this case, how to cluster the edge devices and how to determine cluster heads are important problems.

\subsection{Air-FEEL with New Energy Technologies}

AI model training is generally energy consuming, and the resultant energy cost and carbon emission are becoming serious problems. While conventionally resource management solutions are implemented at the energy demand side, exploiting new energy technologies such as energy harvesting (from, e.g., solar, wind, and ambient radio signals) and smart grids at the energy supply side is becoming a viable new solution to resolve the above problem, as shown in the upper-left subfigure in Fig. \ref{fig:Hierachical}. However, the energy harvesting processes are random and intermittent in nature and the smart grids enable various new features such as dynamic energy pricing and two-way energy trading and sharing. Sophisticated joint energy and Air-FEEL design is desirable for future research.

In addition, {\it wireless power transfer} (WPT) is another powering technique over wireless networks, which can be utilized for the edge server to wirelessly charge edge devices to enable their sustainable and even battery-free operation. 
For such wireless powered Air-FEEL systems, how to jointly optimize the WPT transmission and the Air-FEEL operation is a new technical issue.


\subsection{Air-FEEL with Integrated Sensing and Communications (ISAC) }
ISAC has been recognized as an enabling 6G technology at the physical layer to enable wireless sensing over cellular networks for supporting new intelligent applications. ISAC integrates both functions of data collection and transmission, and its combination with Air-FEEL can be exploited to provide an ultra-low-latency “sensing-communication computation” solution for edge intelligence. For example, as shown in the upper-right subfigure in Fig. \ref{fig:Hierachical}, distributed edge devices can simultaneously transmit probing signals to detect the target and data symbols to the edge server for update aggregation via AirComp. However, the implementation of integrated Air-FEEL and ISAC also faces many challenges. First, since such networks need to support wireless sensing and update aggregation at the same time, the already scarce spectrum resources may become even more strained. Next, different types of edge devices generate different data modalities, and thus how to apply multi-modal sensing data to improve the performance of Air-FEEL deserves more in-depth investigation.


\section{Conclusion}

This article presented an overview on Air-FEEL that exploits the advanced over-the-air data aggregation or AirComp  to enhance the communication efficiency for FEEL towards edge intelligence. We discussed the basic principles, design challenges, advanced techniques, and various new research issues of Air-FEEL. It is our hope that this article can provide new insights on this interesting research topic, and motivate more interdisciplinary research from the communities of wireless communications, machine learning, and computing.

\bibliography{AirCompforFL}
\bibliographystyle{IEEEtran}

\end{document}